\documentclass[a4paper, 10pt]{article}
\usepackage[top=2cm,left=3.5cm,right=3.5cm,bottom=4cm]{geometry}
\usepackage{amssymb}
\usepackage{amsmath}
\usepackage{graphicx}
\usepackage{subfigure}
\usepackage{comment}
\usepackage{cite}
\usepackage[mathscr]{euscript}
\usepackage{multirow}
\begin{document}
\title{\bf Generating Intrinsic Dipole Anisotropy in the Large Scale Structures}
\author{Shamik Ghosh\thanks{shamik@iitk.ac.in} \\ Department of Physics, Indian Institute of Technology, Kanpur,\\ Kanpur - 208016, India}

\maketitle

\begin{abstract}
There have been recent reports of unexpectedly large velocity dipole in the NRAO VLA Sky Survey data. We investigate whether the excess in the NVSS dipole reported can be of cosmological origin. We assume a long wavelength inhomogeneous scalar perturbation of the form $\alpha \sin (\kappa z)$ and study its effects on the matter density contrasts. Assuming an ideal fluid model we calculate, in the linear regime, the contribution of the inhomogeneous mode to the density contrast. We calculate the expected dipole in the LSS for two cases, first assuming that the mode is still superhorizon everywhere, and second assuming the mode is subhorizon, but has crossed the horizon deep in matter domination and is subhorizon everywhere in the region of the survey (NVSS). In both cases we find that such an inhomogeneous scalar perturbation is sufficient to generate the reported values of dipole anisotropy in LSS. For the superhorizon modes we find values which are consistent with both CMB and NVSS results. We also predict signatures for the model which can be tested by future observations.
\end{abstract}
\section{Introduction}
The Cosmological principle assumes that the universe is homogeneous and isotropic on large distance scales. It asserts that there is no special direction in the universe. This assumption leads to the Standard Big Bang Cosmology. But with the modern era of precision cosmology, evidence has been collecting indicating towards deviation from this assumption. 

The data of the Cosmic Microwave Background (CMB) has several features \cite{Tegmark:2004, Eriksen:2004, Ralston:2004, Land:2005, Kim:2010, Samal:2009}, like the CMB dipole power modulation \cite{Ade:2013XXIII}, the alignment of the axes of the quadrupole and octupole etc., which are inconsistent with the standard Hot Big Bang model. The WMAP team had argued that the choice of statistics in the analysis for these anomalies might have led to an overestimation of their the significance \cite{Bennett:2011}. However, the presence of anomalies have been confirmed by the Planck team \cite{Ade:2013XXIII}. Planck have reported \cite{Ade:2013XXIII} a dipole modulation of the power in the range $l=2-600$ at $1.1-3.5 \sigma$, also the octupole-quadrupole alignment has been reported at roughly $98 \%$ level and presence of a cold spot at $(l=32^{\circ} , b=-8^{\circ})$ has been confirmed, while a power deficit at low-$l$ for $20\le l\le 40$ have been reported at $2-2.5 \sigma$.

However, the CMB results are neither the first nor the only evidence of the violation of the Cosmological Principle. Radio polarizations from radio galaxies show large scale dipole pattern \cite{Jain:1998r}. The optical polarization of quasars also show alignment over length scales of Gpc \cite{Hutsemekers:1998, Hutsemekers:2000fv, Jain:2004}. Interestingly, the quadrupole-octupole alignment axis \cite{Tegmark:2004}, the radio polarization dipole axis \cite{Jain:1998r}, the two point correlation in the optical polarization data  \cite{Jain:2004} all align quite close to the CMB dipole axis \cite{Ralston:2004}. The direction is roughly along the direction of the Virgo cluster.

Recently, Singal \cite{Singal:2011}, Gibelyou and Huterer \cite{Gibelyou:2012}, Kothari et. al. \cite{Kothari:2013} and Rubart and Schwarz\cite{Rubart:2013} have reported a dipole anisotropy in the NRAO VLA Sky Survey (NVSS) data, that is much larger than the value of the expected velocity dipole. The Blake and Wall result of 2002 \cite{Blake:2002}, for the NVSS velocity dipole found moderate agreement with the expected value. The newer results \cite{Singal:2011, Gibelyou:2012, Kothari:2013, Rubart:2013} shows little agreement with the expected magnitude of the dipole amplitude but has better agreement with the dipole direction. The NVSS dipole has a direction which is close to the CMB velocity dipole axis.

While the new NVSS dipole results have not been widely accepted, it would be interesting to investigate if the dipole magnitude, as reported, can be of cosmological origin. The objective of this work is to see if a long wavelength  inhomogeneous mode of scalar perturbation can generate an observable dipolar anisotropy in the Large Scale Structure.

In their 2008 paper, Erickcek et al. \cite{Erickcek:2008} used a two field (inflaton and curvaton) model of inflation to generate the dipolar power asymmetry in the CMB data. Here we would use a potential that can be derived from the form of the superhorizon potential they had used and try to see its effects on the matter density contrast. We would work with the inflaton potential only.

\section{Potential}
We are working with a metric of the form
\begin{equation}
ds^2=a(\eta)^2\left[(1+2\Psi(\vec{x},\eta))d\eta^2 -(1+2\Phi(\vec{x},\eta))(dx^2+dy^2+dz^2)\right]
\end{equation}
and with Newtonian gauge we get $\Psi=-\Phi$. The Newtonian potential is given by 
\begin{equation}
\Psi(\vec{x},\eta)=\Psi^{I}(\vec{x},\eta)+\Psi^{A}(\vec{x},\eta)
\end{equation}
Here $\Psi^{I}$ is an isotropic potential while $\Psi^{A}$ is the long wavelength inhomogeneous potential. This inhomogeneous potential is generalized as a sinusoidal potential that would be generated from a primordial sinusoidal fluctuation. The form that we consider for the initial condition is 
\begin{equation}
 \Psi^{A}_i(\vec{x},\eta_i)=\alpha \sin(\kappa z)
\end{equation}
This form can be derived from the generalized sinusoidal potential generated by the inflaton field, assumed by Erickcek et al. \cite{Erickcek:2008}. Their form
\begin{equation}
\label{Eqn1}
\Phi(\vec{x})=\Phi_{\vec{k}}\sin(\vec{k}\cdot\vec{x}+\omega)
\end{equation}
with their assumptions $\hat{k}$ is along $\hat{z}$ and the phase $\omega=0$ reduces to our potential. We identify $\Phi_{\vec{k}}$ with $\alpha$.

The Fourier transform of $\Psi^{A}_i(\vec{x},\eta_i)$ is
\begin{equation}
\label{Eqn16}
\Psi^{A}_i(\vec{k},\eta_i)=\frac{\alpha}{2i}\left[\delta^3(\vec{k}-\kappa\hat{z})-\delta^3(\vec{k}+\kappa\hat{z})\right]
\end{equation}
This is similar in form to the Newtonian potential due to quintessence model constructed by Gordon et. al. \cite{Gordon:2005}.

We assume that the inhomogeneous potential is a single long wavelength mode that remains superhorizon even deep in matter domination. The comoving momentum $k$ is related to the physical momentum as $q(\eta)=k/ a(\eta)$. The condition for horizon crossing is given by $q(\eta) \sim H(\eta)$. This implies the condition that the physical wavelength of the mode is the order of the Hubble size.

\section{Equations}
In our theory we consider density perturbations in an ideal fluid due to the scalar perturbations on our metric. The energy momentum tensor is written as \cite{Gorbunov:2011}
\begin{equation}
T^\mu _\nu =(\hat{\rho}+\hat{p})u^\mu u_\nu - \delta ^\mu _\nu \hat{p}
\end{equation}
where $\hat{\rho}=\rho+\delta \rho$ and $\hat{p}=p+\delta p$. Also $u^\mu$ is the four velocity, containing both a background part and a small perturbation. We assume that the fluid has no bulk velocity and hence the only non-zero component of the background is $u^0$. We would write $u^0=a^{-1}(1+\delta u^0)$ and $u^i=a^{-1}v^i$.

The linearised Einstein's equation when written in Fourier space we can separate out the different $k$ modes when working in the linear theory. On doing the Fourier transform every derivative would pull down a $ik$. They finally give the form
\begin{align}
\label{Eqn4}
k ^2 \Psi +3 \frac{a'}{a} \Psi'+3\frac{a'^2}{a^2} \Psi  &= -4 \pi G a^2 \sum _\lambda \delta \rho _{\lambda}\\
\label{Eqn5}
\Psi'+\frac{a'}{a} \Psi &= -4 \pi G a^2\sum _\lambda[(\rho+p)v]_{\lambda}\\
\label{Eqn6}
\Psi''+3 \frac{a'}{a} \Psi'+\left(2\frac{a''}{a}-\frac{a'^2}{a^2}\right)\Psi &= 4 \pi G a^2 \sum _\lambda\delta p _{\lambda}
\end{align}

We can define the relative perturbations in the energy density for the $\lambda$ component for $k$th mode as $\delta _\lambda = \delta \rho _{\lambda} /\rho _{\lambda}$. Then using $p_\lambda=w_\lambda \rho_\lambda$
and $\delta p _\lambda = u_{\lambda} ^2\delta \rho_\lambda$, the first order conservation equations are as follows
\begin{align}
\label{Eqn7}
\delta'_\lambda+3\frac{a'}{a}(u_{\lambda}^2-w_\lambda)\delta_\lambda -(1+w_\lambda)k^2v_\lambda &=3(1+w_\lambda)\Psi'\\
\label{Eqn8}
[(1+w_\lambda)v_\lambda]'+\frac{a'}{a}(1-3w_\lambda)(1+w_\lambda)v_\lambda
+u_{\lambda}^2\delta_\lambda &=-(1+w_\lambda)\Psi
\end{align}

Using $\delta p _\lambda = u_{\lambda} ^2\delta \rho_\lambda$ and the Friedmann Equations, we arrive at the differential equation for the time evolution of $\Psi$
\begin{equation}
\label{Eqn9}
\Psi''+3\frac{a'}{a}(1+u^2)\Psi'+u^2k^2\Psi=0
\end{equation}

With Eqns(\ref{Eqn4}-\ref{Eqn9}) defined we are going to study the behaviour of the density perturbations.

\section{Matter Density Perturbation}
Since we are working in the linear regime, we can solve for the inhomogeneous long wavelength mode separately and add the result to the standard results for homogeneous perturbations. We solve the Eqns(\ref{Eqn4}-\ref{Eqn9}) by use of standard techniques \cite{Gorbunov:2011}. The results for the long wavelength mode being both subhorizon and superhorizon during present time are standard results of the text. We will discuss them below and then proceed to calculate the value of the dipole in either cases.

\subsection{Mode Superhorizon}
For a superhorizon mode one can make the assumption that $k \rightarrow 0$, hence dropping all the terms containing $k$ from the equations. While $u^2=1/3$ during radiation domination, it is zero during matter domination. The solutions for superhorizon mode potentials of Eqn(\ref{Eqn9}) shows that the potential is effectively constant and equal to the initial value at each epoch. We calculate the values of the
density contrast in different component by making use of the curvature perturbation, $\zeta(\vec{r})$, which remains constant.

In deep matter domination the density perturbations for the superhorizon mode is given by
\begin{align}
\label{Eqn10}
\delta_B^A(k)=\delta_{CDM}^A(k) &=-\frac{9}{5}\Psi^A_i(k)\\
\label{Eqn11}
\delta_{\gamma}^A(k) &=-\frac{12}{5}\Psi^A_i(k)
\end{align}
With $u^2 = w =0$ in matter dominated epoch, $k\rightarrow 0$ for superhorizon mode and $\Psi ' =0$ since the potential is constant, from Eqn(\ref{Eqn7}) we find $\delta '_\lambda =0$. Thus for superhorizon modes the density contrast for every component remains constant and equal to the initial value set at the beginning of the epoch.

For our current epoch which is not wholly dominated by matter, the contribution of the dark energy needs to be taken into consideration. We account for the presence of dark energy by introducing a suppression factor $g(\eta)$, \cite{Gorbunov:2011}. The factor, $g(\eta)$, is a number corresponding to the present conformal time, equal to 0.76 at present. Thus with the dark energy suppression considered, the matter density perturbation of a superhorizon mode is given by
\begin{equation}
\label{Eqn12}
\delta_B^A(k)=\delta_{CDM}^A(k) =-\frac{9}{5}g(\eta)\Psi^A_i(k)
\end{equation}
where $g(\eta)$ is a numerical factor. 

This is the result for the inhomogeneous superhorizon mode only. The density perturbations arising from the homogeneous and isotropic modes can be simply added to the above result since we are working in linear regime. The $\Psi^A_i(k)$ in Eqn(\ref{Eqn12}) is given by the expression in Eqn(\ref{Eqn16}). After taking a Fourier transform of the total matter density perturbation it becomes
\begin{equation}
\label{Eqn17}
\delta_M(\vec{r},\eta) \approx -\left[\widetilde{F}^I(\vec{r})\eta^2+\frac{9}{5}\alpha \sin(\kappa z)\right]g(\eta)
\end{equation} 
considering that the isotropic and homogeneous modes are already subhorizon. The exact form of the $\widetilde{F}^I(\vec{r})$ is can be calculated from standard isotropic results available in texts \cite{Gorbunov:2011}.
\subsection{Mode Subhorizon}
If the mode has gone subhorizon, the terms containing $k$ cannot be neglected. Since we are considering a long wavelength mode, it would cross the horizon deep in matter domination. For such mode Eqn(\ref{Eqn9}), with $u^2 = 0$ in matter domination, would give a solution for the time variation of the potential that would be mostly constant and equal to the value at horizon crossing.

The initial values for the density perturbations are set by Eqns(\ref{Eqn10}-\ref{Eqn11}). The time evolution of the density perturbations for each component from Eqns(\ref{Eqn7}-\ref{Eqn8}) with relevant values of the constants for the respective component in the matter dominated epoch and also noting that the potential has zero time derivative. The solution for the matter density perturbations, with the dark energy suppression factor for late times included, are given by
\begin{equation}
\label{Eqn15}
\delta_{CDM}^A (k,\eta)=\delta_B^A(k, \eta) =-g(\eta)\left[\frac{9}{5}\Psi^A_i(k)+\frac{1}{6}k^2(\eta-\eta_\times)^2\Psi^A_i(k)\right]
\end{equation}
where $\eta_\times$ is the conformal time of horizon crossing. Adding the contribution from the isotropic and homogeneous modes and taking Fourier transform the result becomes
\begin{equation}
\label{Eqn18}
\delta_M(\vec{r},\eta) \approx -\left[\widetilde{F}^I(\vec{r})\eta^2+\left\{\frac{9}{5}+\frac{1}{6}\kappa^2(\eta-\eta_\times)^2\right\}\alpha \sin(\kappa z)\right]g(\eta)
\end{equation}
Like the result in the previous section here the subscript $M$ denotes either baryons or CDM.

\section{Dipole Calculations}
To calculate the dipole from the density contrast we need to expand the density contrast $\delta(\vec{r}, \eta)$ in spherical harmonics and then calculate the projected dipole by integrating over $z$. We would compare this projected dipole term with observational results to set limits on the amplitude.

\subsection{Expansion Coefficient Calculations}
By our initial choice of our coordinates we have assumed the inhomogeneity along the $z$ direction. We would work with only the inhomogeneous term in the expression for $\delta_M (\vec{r}, \eta)$ because the isotropic and homogeneous term in the expression, $\widetilde{F}^I(\vec{r})$, would not contribute to the dipole or higher order terms when expanded in spherical harmonics. Hence we can write the following expression
\begin{equation}
\delta(r,\theta, \phi)=\sum_{l=0}^{\infty}\sum_{m=-l}^{m=+l}a_{lm}(r)Y_{lm}(\theta, \phi).
\label{Eqn19}
\end{equation}
We expand the density contrast $\delta^A(r, \theta, \eta)$ in spherical harmonics as in Eqn(\ref{Eqn19}), where the coefficients $a_{lm}(r,\eta)$ are given by
\begin{equation}
a_{lm}=\int \delta^A(r, \theta, \eta)Y^* _{lm}(\theta, \phi)d\Omega.
\end{equation}

\subsubsection{Mode Superhorizon}
The density contrast while considering a superhorizon perturbation mode is given in Eqn(\ref{Eqn17}). The first term in the bracket would not contribute to the dipole calculations since it is homogeneous and isotropic by assumption. We will only consider the second term in the expression for the calculation. Hence for a superhorizon mode
\begin{align}
\delta_M^A(\vec{r}) &=- \frac{9}{5}g(\eta)\alpha \sin (\kappa z) \nonumber \\
\label{Eqn21}
                    &=-\frac{9}{5}g(\eta)\alpha \sin (\kappa r \cos \theta)
\end{align}
Using Eqn(\ref{Eqn21}) we calculate the $a_{lm}$s for $l=0$,$1$. Except for $a_{10}$ all others are zero. The expression for $a_{10}$ is
\begin{equation}
\label{Eqn22}
a_{10}(r)=\frac{18}{5}\sqrt{3\pi}g(\eta) \alpha \left[\frac{\cos(\kappa r)}{\kappa r}-\frac{\sin (\kappa r)}{\kappa^2 r^2}\right]
\end{equation}
Note that this is still a function of $r$. We would require to integrate out $r$ to get the result for the projected dipole.

\subsubsection{Mode Subhorizon}
The subhorizon mode density contrast is given by Eqn(\ref{Eqn18}) and like in the case for superhorizon modes the contribution from the first term in the bracket, to the dipole is zero since we have assumed it to be isotropic and homogeneous. The density contrast for the subhorizon mode is
\begin{align}
\delta_M^A(\vec{r}) &=-\left\{\frac{9}{5}+\frac{1}{6}\kappa^2(\eta-\eta_\times)^2\right\}g(\eta)\alpha \sin(\kappa z) \nonumber \\
\label{Eqn23}
                    &=-\left\{\frac{9}{5}+\frac{1}{6}\kappa^2(\eta-\eta_\times)^2\right\}g(\eta)\alpha \sin(\kappa r \cos \theta)
\end{align}
With Eqn(\ref{Eqn23}) we again calculate the $a_{lm}$s for $l=0$, $1$ and except for $a_{10}$ all others are zero. the form of $a_{10}$ works out to be
\begin{equation}
\label{Eqn24}
a_{10}(r)=\left\{\frac{9}{5}+\frac{1}{6}\kappa^2(\eta-\eta_\times)^2\right\}\sqrt{3\pi}g(\eta) 2 \alpha \left[\frac{\cos(\kappa r)}{\kappa r}-\frac{\sin (\kappa r)}{\kappa^2 r^2}\right]
\end{equation}
which is again dependent on $r$.

\subsection{Projected Dipole}
Both the relations Eqn(\ref{Eqn22}) and Eqn(\ref{Eqn24}), for the superhorizon and subhorizon mode $a_{10}$, are still functions of the redshift $z$ since they are $r$ and $\eta$ dependent. We replace $r$ and $\eta$ by these relations
\begin{align}
r(z)=\frac{1}{a_0H_0}\int_{0}^{z}\frac{dz'}{\sqrt{\Omega_{\Lambda}+\Omega_M(1+z')^3}} && \eta(z)=\frac{1}{a_0H_0}\int_{z}^{\infty}\frac{dz'}{\sqrt{\Omega_{\Lambda}+\Omega_M(1+z')^3}} \nonumber
\end{align}
in terms of redshift and integrate over the volume of observation. Since the NVSS catalogue stretches to a redshift of 2, we will integrated the expressions of $a_{10}$ form $z=0$ to $z=2$. This is equivalent to projecting the density contrast between $z=0$ to $z=2$ onto a sphere at $z=2$. This removes the distance information in $a_{10}$.

The integration by redshift cannot be performed analytically since there are no closed form expressions for $r$ and $\eta$. We perform the integral numerically with the standard Simpson's 1/3 Rule implemented with a C++ routine. Since we have two unknown variables $k$ and $\alpha$, we perform the integrations for different fixed values of $k$, both for the superhorizon and the subhorizon modes. After integration we would get for each fixed value of $k$ a numerical result multiplying $\alpha$. For superhorizon modes the results are tabulated in Table \ref{Tbl1}.
\begin{table}[h]
\centering
\begin{tabular}{|c|c|}
\hline
$k$ (in units of $H_0$) & Numerical result $\bar{a}_{10}$\\
\hline
0.10 &	0.41$\alpha$\\
0.09 &	0.36$\alpha$\\
0.08 &	0.33$\alpha$\\
0.07 &	0.28$\alpha$\\
0.06 &	0.24$\alpha$\\
0.05 &	0.20$\alpha$\\
0.04 &	0.16$\alpha$\\
0.03 &	0.12$\alpha$\\
\hline
\end{tabular}
\caption{Superhorizon Mode Integral Results}
\label{Tbl1}
\end{table}

\begin{table}[h]
\centering
\begin{tabular}{|c|c|c|}
\hline
$z$ of Re-entry & $k$ (in units of $H_0$) & Numerical result $\bar{a}_{10}$\\
\hline
3 &	1.08 &	4.13$\alpha$\\
4 &	1.19 &	4.66$\alpha$\\
5 &	1.30 &	5.22$\alpha$\\
6 &	1.40 &	5.81$\alpha$\\
7 &	1.50 &	6.42$\alpha$\\
8 &	1.59 &	7.04$\alpha$\\
\hline
\end{tabular}
\caption{Subhorizon Mode Integral Results}
\label{Tbl2}
\end{table} 
For subhorizon modes we are careful not to include values of $k$ which re-enter the horizon very close to today and again re-exit. The condition for re-entry is given by 
\begin{equation}
\frac{k}{a(z)}\sim H_0\sqrt{\Omega_\Lambda + \Omega_M(1+z)^3}.
\end{equation}
The modes smaller than $a(z)H(z)$ are superhorizon while those larger are subhorizon. Modes which cross the horizon after $z\sim 3$ re-exit the horizon before $z=0$. We have avoided these modes for the subhorizon calculations. Given in Table \ref{Tbl2} are the numerical integration results with the value of $k$ in units of $H_0$ corresponding to the redshift horizon re-entry.

\section{Fitting the Experimental Results}
Following the original Blake and Wall results \cite{Blake:2002} there has been several papers \cite{Singal:2011, Gibelyou:2012, Kothari:2013, Rubart:2013} all of which have claimed larger values for the dipole amplitude than what is typically expected. Since we are interested in studying the intrinsic dipole in the large scale structure it is important to note that there are other possible sources for a dipole anisotropy in the LSS. We shall investigate them to account for them in our calculations.

\textit{Kinematic Dipole}: The primary reason for a dipole to be found in the LSS density field is of kinematic origin, just like the CMB kinematic dipole. It is important to realize that the frame on the earth is in relative motion with the CMB frame. This is generated since the Earth goes around the Sun at $\sim 30$ km/s, the Sun moves with respect to the Local Group at $\sim 306$ km/s while the Local Group moves at $\sim 622$ km/s with respect to the CMB. The resultant motion gives rise to an overall velocity of $369.0 \pm 0.9$ km/s along $(l,b)=(263.99,48.26)\pm(0.14,0.03)$ in galactic coordinates \cite{Hinshaw:2008}. This effective motion of our observation frame results in a kinematic dipole in the CMB. 

The motion of observation frame would generate a dipolar modulation of the \emph{observed} density field by two physical process, (1) relativistic aberration and (2) Doppler effect. 

Due to effective motion of the observer's frame there would be a significant relativistic aberration. The effect is of the order of $v/c=1.23 \times 10^{-3}$.  Due to the motion of the frame along the $z$ direction (say), the azimuthal angles in the observer's frame would be same as the CMB frame while the angle $\theta$ with the $z$ axis would change. This implies that objects located in one hemisphere are partly shifted towards the other hemisphere in the direction of motion. The number of objects being conserved, this results in a larger number density in one hemisphere compared to the other. Thus the density field would show dipolar modulation due to relativistic aberration.

Every survey works in a particular frequency band. Sources in the direction of motion get blue shifted and certain sources with frequencies below the frequency band gets blue shifted into the observation. While in the other hemisphere, in the direction opposite to the motion, sources are red shifted out of the frequency range, thereby resulting in an effective dipole in the density field measured by the survey. 

 We can model the flux density of radio sources $S$ as $S \propto \nu ^{-p}$ while the number density of radio sources with flux density greater than a certain limiting flux is given by $N(S>S_{lim})=S^{-x}$. Then the total kinematic dipole amplitude due to the combination of both the effects is given by \cite{Ellis:1984}
 \begin{equation}
 D_{kin}=\frac{v}{c}\left[2+x(1+p)\right].
 \end{equation}
Now with $v/c=1.23 \times 10^{-3}$, $x \approx 1$, $p\sim 0.75$ gives the theoretical prediction for the kinematic dipole in the NVSS as \cite{Gibelyou:2012}
\begin{equation}
D_{kin}=0.0046 \pm 0.0029
\end{equation}

In the results being considered here \cite{Singal:2011, Gibelyou:2012, Kothari:2013, Rubart:2013} the authors have compared their results to the kinematic dipole and found to have exceeded the estimates for the kinematic dipole. Since we are interested in an intrinsic dipole, which might be present over and above the kinematic dipole, we must subtract the predicted value of the kinematic dipole form the results reported in these papers. The results reported in the papers \cite{Singal:2011, Gibelyou:2012, Kothari:2013, Rubart:2013} and the kinematic dipole subtracted residual are given in Table \ref{Tbl3}.

\begin{table}[h]
\centering
\begin{tabular}{|p{3cm}|c|c|}
\hline
Author & Dipole Amplitude  $D_o$ & $D_{res}=D_o-D_{kin}$\\
\hline
Kothari et. al. \cite{Kothari:2013} (Number Count) & $0.0151 \pm 0.0030$ & $0.010 \pm 0.004$\\ 
\hline
Kothari et. al. \cite{Kothari:2013} (Sky Brightness) & $0.0166 \pm 0.0031$ & $0.012 \pm 0.004$\\
\hline
Singal \cite{Singal:2011} & $0.019 \pm 0.004$ & $0.014\pm 0.005$\\
\hline
Gibelyou \& Heutrer \cite{Gibelyou:2012} & $0.027\pm 0.005$ & $0.022 \pm 0.006$\\
\hline
Rubart \& Schwarz \cite{Rubart:2013} & $0.018 \pm 0.006$ & $0.013 \pm 0.007$\\
\hline
\end{tabular}
\caption{Total and Residual Dipole Amplitudes}
\label{Tbl3}
\end{table}

We would proceed to model the dipole in a way similar to the treatment of Gibelyou and Huterer \cite{Gibelyou:2012}. We consider a dipole modulation in the sky along direction $\hat{\kappa}$ with a magnitude $D$. Then the observed density field $N$ can be written as the following function of direction in the sky $\hat{n}$
\begin{equation}
N(\hat{n})\approx \left[1+D(\hat{\kappa}\centerdot\hat{n})\right]\bar{N}
\end{equation} 
where $\bar{N}$ is the intrinsic isotropic field. Hence the contrast in the density field defined as above is given by
\begin{equation}
\frac{\delta N}{\bar{N}}(\hat{n})\approx D(\hat{\kappa}\centerdot\hat{n})
\end{equation} 

Now to compare the dipole amplitude with the dipole power term in the expansion of $\delta$, we assume without loss of generality that the dipolar modulation direction, $\kappa$ is along the $z$ direction. Then we can write
\begin{equation}
\frac{\delta N}{\bar{N}}(\hat{n})\approx D(\hat{\kappa}\centerdot\hat{n})=D \cos \theta = D\sqrt{\frac{4 \pi}{3}}Y_{10}(\hat{n}).
\end{equation} 
Comparing with Eqn(\ref{Eqn19}) one can write
\begin{equation}
\label{Eqn25}
\bar{a}_{10}=D\sqrt{\frac{4 \pi}{3}}
\end{equation}
where $\bar{a}_{10}$ is the redshift integrated $a_{10}(z)$ and $D$ is the residual dipole amplitude $D_{res}$ after subtraction of the kinematic component.

\begin{table}[h]
\centering
\begin{tabular}{|p{1cm}|p{2cm}|p{2cm}|p{2cm}|p{2cm}|p{2cm}|}	
\hline
k (in units of $H_0$) & Kothari et. al. (Number) & Kothari et. al. (Sky) & Singal & Gibelyou \& Heutrer & Rubart \& Schwarz \\ 
\hline
0.10 &	$0.05 \pm	0.02$ &	$0.06 \pm	0.02$ &	$0.07 \pm	0.03$ &	$0.11 \pm	0.03$ & $0.07 \pm 0.04$\\
0.09 &	$0.06 \pm	0.02$ &	$0.07 \pm	0.02$ &	$0.08 \pm	0.03$ &	$0.12 \pm	0.03$ & $0.07 \pm	0.04$\\
0.08 &	$0.06 \pm	0.03$ &	$0.08 \pm	0.03$ &	$0.09 \pm	0.03$ &	$0.14 \pm	0.04$ & $0.08 \pm	0.04$\\
0.07 &	$0.07 \pm	0.03$ &	$0.09 \pm	0.03$ &	$0.10 \pm	0.04$ &	$0.16 \pm	0.04$ & $0.09 \pm	0.05$\\
0.06 &	$0.08 \pm	0.03$ &	$0.10 \pm	0.03$ &	$0.12 \pm	0.04$ &	$0.18 \pm	0.05$ & $0.11 \pm	0.06$\\
0.05 &	$0.10 \pm	0.04$ &	$0.12 \pm	0.04$ &	$0.14 \pm	0.05$ &	$0.22 \pm	0.06$ & $0.13 \pm	0.07$\\
0.04 &	$0.13 \pm	0.05$ &	$0.15 \pm	0.05$ &	$0.18 \pm	0.06$ &	$0.28 \pm	0.08$ & $0.16 \pm	0.09$\\
0.03 &	$0.17 \pm	0.07$ &	$0.20 \pm	0.07$ &	$0.24 \pm	0.08$ &	$0.37 \pm	0.10$ & $0.22 \pm	0.12$\\
\hline
\end{tabular}
\caption{Value of $\alpha$ for superhorizon mode for different values of $k$ fitting to the four different amplitude results}
\label{Tbl4}
\end{table}
Using Eqn(\ref{Eqn25}) with data from Table \ref{Tbl1} and Table \ref{Tbl3} we calculate the values of $\alpha$ for different superhorizon modes and they are tabulated in Table \ref{Tbl4}. Similar calculations with Table \ref{Tbl2} gives the values for subhorizon modes which are tabulated in Table \ref{Tbl5}.

\begin{table}[h]
\centering
\begin{tabular}{|p{1cm}|p{2cm}|p{2cm}|p{2cm}|p{2cm}|p{2cm}|}	
\hline
k (in units of $H_0$) & Kothari et. al. (Number) & Kothari et. al. (Sky) & Singal & Gibelyou \& Heutrer & Rubart \& Schwarz \\ 
\hline
1.079 &	$0.005 \pm	0.002$ &	$0.006 \pm	0.002$ &	$0.007 \pm	0.002$ &	$0.011 \pm	0.003$ & $0.006 \pm	0.003$\\
1.195 &	$0.004 \pm	0.002$ &	$0.005 \pm	0.002$ &	$0.006 \pm	0.002$ &	$0.010 \pm	0.003$ & $0.006 \pm	0.003$\\
1.304 &	$0.004 \pm	0.002$ &	$0.005 \pm	0.002$ &	$0.005 \pm	0.002$ &	$0.009 \pm	0.002$ & $0.005 \pm	0.003$\\
1.405 &	$0.004 \pm	0.001$ &	$0.004 \pm	0.001$ &	$0.005 \pm	0.002$ &	$0.008 \pm	0.002$ & $0.005 \pm	0.002$\\
1.500 &	$0.003 \pm	0.001$ &	$0.004 \pm	0.001$ &	$0.004 \pm	0.002$ &	$0.007 \pm	0.002$ & $0.004 \pm	0.002$\\
1.590 &	$0.003 \pm	0.001$ &	$0.003 \pm	0.001$ &	$0.004 \pm	0.001$ &	$0.006 \pm	0.002$ & $0.004 \pm	0.002$\\
\hline
\end{tabular}
\caption{Value of $\alpha$ for subhorizon mode for different values of $k$ fitting to the four different amplitude results}
\label{Tbl5}
\end{table} 

\begin{figure}[h]
\centering
\includegraphics[scale=0.5]{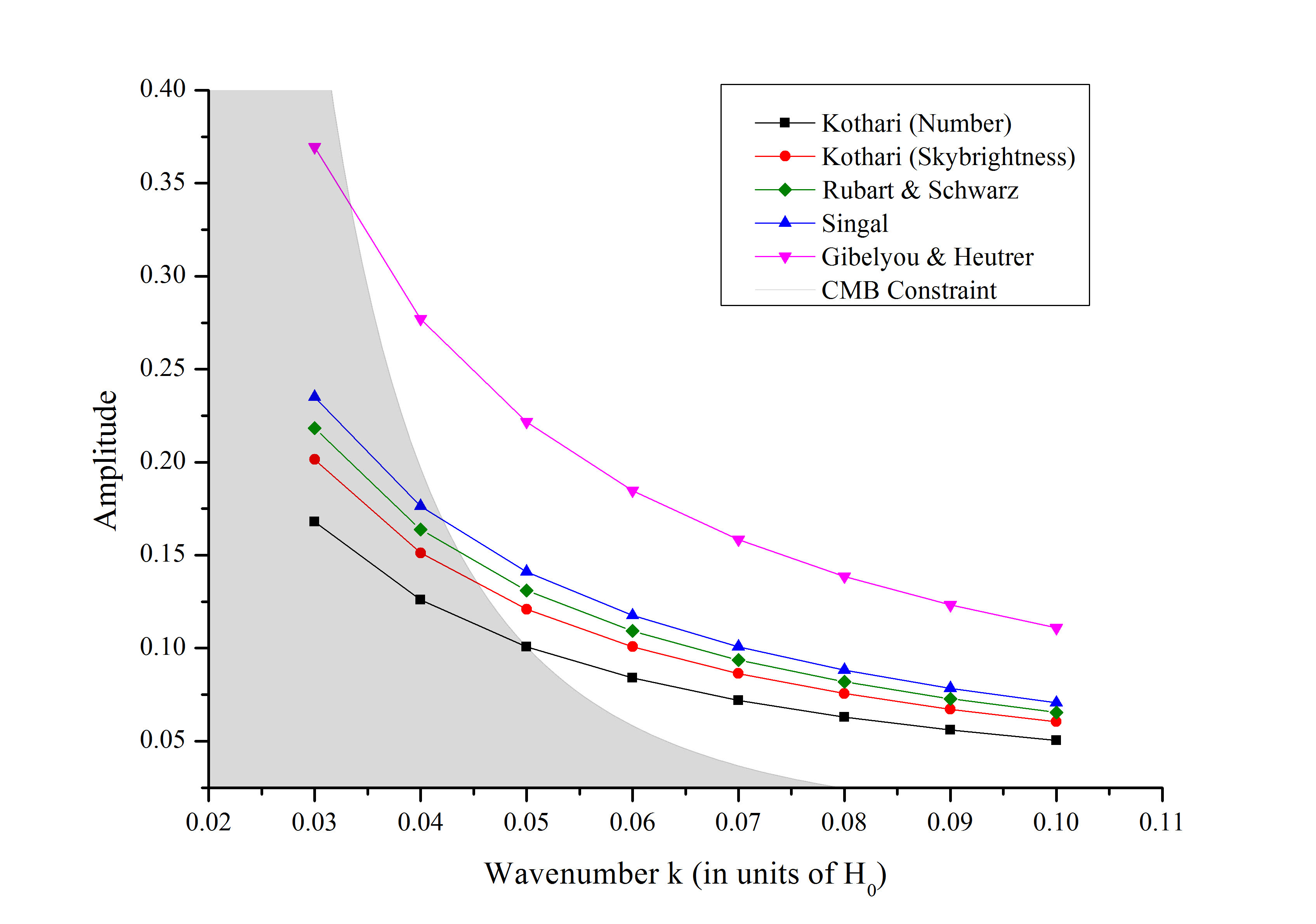}
\caption{Plot of $\alpha$ versus $k$ for superhorizon modes with CMB constraint equation. The points within the grey region is allowed.}
\label{Fig1}
\end{figure}

Also note that for superhorizon modes there is a constraint from the CMB data. This constraint from the CMB quadrupole and octupole is calculated by Erickcek et. al. \cite{Erickcek:2008}. It can be rewritten as
\begin{equation}
\label{Eqn26}
k^3\alpha \le 1.26 \times 10^{-5}H_0^3
\end{equation}
We plot Eqn(\ref{Eqn26}) and the results of Table \ref{Tbl4} in Fig.\ref{Fig1}. The points on the curves that lie inside the shaded region are consistent with both the CMB and NVSS observations.

There should also be CMB constraints for the subhorizon modes arising from the SW and ISW effects of the potential. But we have not calculated these constraints. These should provide further constraints on the choice of $k$ and $\alpha$. It should also be noted that there is considerable error in the results for $\alpha$. For the superhorizon modes the amplitude of the perturbation is of the order of that of the NVSS dipole amplitude.

Here we would also like to make a comment on the robustness of the model to the addition of a constant phase shift of $\omega$ as for the potential assumed by Erickcek et. al. \cite{Erickcek:2008}, previously mentioned in Eqn(\ref{Eqn1}). If $\omega \neq 0$, we would have an additional $\cos \omega$ term multiplying our result for $a_{10}(r)$, in Eqns(\ref{Eqn22}, \ref{Eqn24}). This surmounts to multiplying our final result with a fraction. The model is fairly robust for nearly all values of $\omega$, if we are considering a subhorizon mode. For superhorizon mode, with very small $\kappa$, the value of $|\cos \omega |$ should not differ by a large extent from unity. If it differs significantly, the amplitude $\alpha$ required to fit the NVSS dipole data would be significantly large. Choosing $\cos \omega \sim 0$ amounts to a phase shift that changes the symmetry of the potential and hence such a mode should not be expected to a generate dipole anisotropy in the matter density contrast.

\section{Future Tests of the Model}
The class of models discussed here as a method of generating anisotropy in the large scale structure have some typical signatures which could be tested in the near future. Since the model generates an inhomogeneity in density contrast, it should lead to an inhomogeneity in the number count distribution of galaxies. To examine this feature of the model we assume a linear galaxy bias to relate the galaxy density contrast $\delta_g$ to the actual matter density contrast as
\begin{equation}
\delta_g(\vec{r}) = \frac{n(\vec{r})-\bar{n}}{\bar{n}} = b_1 \delta(\vec{r})
\end{equation}
where $n(\vec{r})$ is the galaxy number count at $\vec{r}$, $\bar{n}$ is the average number count, $b_1$ is the linear bias constant. This leads to 
\begin{equation}
n(\vec{r})=\bar{n}\left[1 + K\delta(\vec{r})\right]
\end{equation}
where $K = 1/b_1$. The matter density contrast $\delta (\vec{r})$ is given by of Eqn(\ref{Eqn17}) or Eqn(\ref{Eqn18}) and using a general expression to represent both
\begin{equation}
n(\vec{r})=\bar{n}\left[1 - K\left\{\widetilde{F}^I(r)\eta^2 + F^A(\kappa z)\right\}g(\eta)\right]
\end{equation} 
where $F^A(\kappa z) \propto \alpha \sin(\kappa z)$. The proportionality constant would depend on whether the mode is superhorizon or subhorizon. It is easy to see that such class of models which break isotropy with introduction of a sinusoidal mode to introduce inhomogeneity would lead to an inhomogeneous redshift distribution along the preferred direction and it would have a typical sinusoidal variation.

However the introduction of the long wavelength inhomogeneous perturbation should not change the luminosity distribution for the galaxies which depends on local astrophysics which are affected by potentials which are more local than the perturbation. 

While the EUCLID mission would cover only about half the sky, if in future, there is an all sky mission which is deeper than the NVSS, then assuming the model in this paper the dipole for the deeper survey should have a larger magnitude. So another test would be to match the prediction for the minimum expected dipole for a survey depth of $z \sim 3$ with that of future surveys. Using the smallest fit values for $\alpha$ we find that the minimum expected dipole would be $3 \times 10^{-2}$ for subhorizon mode and $4 \times 10^{-2}$ for superhorizon mode. These rather large predictions are due to the nature of the potential. As the survey volume grows the inhomogeneities grow as potential reaches the extrema. It might be possible to test the signatures of the model in the near future.

\section{Conclusions}
Starting from an ansatz for an inhomogeneous scalar perturbation we have been able to show that the reported excesses in the NVSS dipole can be explained as being of cosmological origin. While there are still divided opinion about the reported excess in the results we have shown that such excesses can also be artifacts of departures from cosmological principle. While the error bars on the results are quite large, modes with amplitude of the order of the NVSS dipole, or smaller can generate the dioples reported. For superhorizon modes there exists values of $k$ and $\alpha$ which are also consistent with the CMB constraint. We have pointed out that a model of this class would have some very definite signatures which can be tested in the near future.

\bibliographystyle{unsrt}
\bibliography{Paper}

\begin{thebibliography}{10}

\bibitem{Tegmark:2004}
Ang\'elica de~Oliveira-Costa, Max Tegmark, Matias Zaldarriaga, and Andrew
  Hamilton.
\newblock Significance of the largest scale cmb fluctuations in wmap.
\newblock {\em Phys. Rev. D}, 69:063516, Mar 2004.

\bibitem{Eriksen:2004}
H.K. Eriksen, F.K. Hansen, A.J. Banday, K.M. Gorski, and P.B. Lilje.
\newblock {Asymmetries in the Cosmic Microwave Background anisotropy field}.
\newblock {\em ApJ}, 605:14--20, 2004.

\bibitem{Ralston:2004}
John~P. Ralston and Pankaj Jain.
\newblock {The Virgo alignment puzzle in propagation of radiation on
  cosmological scales}.
\newblock {\em IJMPD}, 13:1857--1878, 2004.

\bibitem{Land:2005}
Kate Land and Joao Magueijo.
\newblock {Is the Universe odd?}
\newblock {\em Phys. Rev. D}, 72:101302, 2005.

\bibitem{Kim:2010}
Jaiseung Kim and Pavel Naselsky.
\newblock {Anomalous parity asymmetry of the Wilkinson Microwave Anisotropy
  Probe power spectrum data at low multipoles}.
\newblock {\em ApJ}, 714:L265--L267, 2010.

\bibitem{Samal:2009}
Pramoda~Kumar Samal, Rajib Saha, Pankaj Jain, and John~P. Ralston.
\newblock {Signals of Statistical Anisotropy in WMAP Foreground-Cleaned Maps}.
\newblock {\em MNRAS}, 396:511, 2009.

\bibitem{Ade:2013XXIII}
P.A.R. Ade et~al.
\newblock {Planck 2013 results. XXIII. Isotropy and Statistics of the CMB}.
\newblock 2013.

\bibitem{Bennett:2011}
C.L. Bennett, R.S. Hill, G.~Hinshaw, D.~Larson, K.M. Smith, et~al.
\newblock {Seven-Year Wilkinson Microwave Anisotropy Probe (WMAP) Observations:
  Are There Cosmic Microwave Background Anomalies?}
\newblock {\em ApJS}, 192:17, 2011.

\bibitem{Jain:1998r}
P.~Jain and J.~P. Ralston.
\newblock Anisotropy in the propagation of radio polarizations from
  cosmologically distant galaxies.
\newblock {\em MPLA}, 14:417--432, 1998.

\bibitem{Hutsemekers:1998}
D.~Hutsem\'{e}kers.
\newblock Evidence for very large-scale coherent orientations of quasar
  polarization vectors.
\newblock {\em A\&A}, 332:410--428, 1998.

\bibitem{Hutsemekers:2000fv}
D.~Hutsem\'{e}kers and H.~Lamy.
\newblock Confirmation of the existence of coherent orientations of quasar
  polarization vectors on cosmological scales.
\newblock {\em A\&A}, 367(2):381--387, 2001.

\bibitem{Jain:2004}
Pankaj Jain, Gaurav Narain, and S~Sarala.
\newblock {Large-scale alignment of optical polarizations from distant QSOs
  using coordinate-invariant statistics}.
\newblock {\em MNRAS}, 347(2):394--402, 2004.

\bibitem{Singal:2011}
Ashok~K. Singal.
\newblock {Large peculiar motion of the solar system from the dipole anisotropy
  in sky brightness due to distant radio sources}.
\newblock {\em ApJL}, 742:L23, 2011.

\bibitem{Gibelyou:2012}
Cameron Gibelyou and Dragan Huterer.
\newblock Dipoles in the sky.
\newblock {\em MNRAS}, 427(3):1994--2021, 2012.

\bibitem{Kothari:2013}
Rahul Kothari, Abhishek Naskar, Prabhakar Tiwari, Sharvari Nadkarni-Ghosh, and
  Pankaj Jain.
\newblock {Dipole anisotropy in flux density and source count distribution in
  radio NVSS data}.
\newblock 2013.

\bibitem{Rubart:2013}
Matthias Rubart and Dominik~J. Schwarz.
\newblock {Cosmic radio dipole from NVSS and WENSS}.
\newblock {\em A \& A}, 555:A117, 2013.

\bibitem{Blake:2002}
Chris Blake and Jasper Wall.
\newblock A velocity dipole in the distribution of radio galaxies.
\newblock {\em Nature}, 416(6877):150--152, 2002.

\bibitem{Erickcek:2008}
Adrienne~L. Erickcek, Marc Kamionkowski, and Sean~M. Carroll.
\newblock A hemispherical power asymmetry from inflation.
\newblock {\em Phys. Rev. D}, 78:123520, Dec 2008.

\bibitem{Gordon:2005}
Christopher Gordon, Wayne Hu, Dragan Huterer, and Thomas~M. Crawford.
\newblock {Spontaneous isotropy breaking: a mechanism for cmb multipole
  alignments}.
\newblock {\em Phys. Rev. D}, 72:103002, 2005.

\bibitem{Gorbunov:2011}
Dmitry~S. Gorbunov and Valery~A. Rubakov.
\newblock {\em Introduction to the theory of the early universe: Cosmological
  Perturbations and Inflationary Theory}.
\newblock World Scientific, 2011.

\bibitem{Hinshaw:2008}
G.~Hinshaw et~al.
\newblock {Five-Year Wilkinson Microwave Anisotropy Probe (WMAP) Observations:
  Data Processing, Sky Maps, and Basic Results}.
\newblock {\em ApJS.}, 180:225--245, 2009.

\bibitem{Ellis:1984}
G.~F.~R. {Ellis} and J.~E. {Baldwin}.
\newblock {On the expected anisotropy of radio source counts}.
\newblock {\em MNRAS}, 206:377--381, January 1984.

\end{thebibliography}
\end{document}